\documentclass[%
 reprint,
 amsmath,amssymb,
 aps,twocolumn
]{revtex4-1}
\usepackage{color}
\usepackage{graphicx}
\usepackage{dcolumn}
\usepackage{bm}
\usepackage[pagebackref=false,colorlinks,linkcolor=blue,citecolor=blue]{hyperref}

\begin{document}

\preprint{APS/123-QED}

\title{Three parameter metrics  in the presence of a scalar field in four and higher dimensions}

\author{Alireza Azizallahi}
   \email{a.azizallahi@ph.iut.ac.ir}
   \author{Behrouz Mirza   }
  \email{b.mirza@iut.ac.ir}
  \author{Arash Hajibarat}
  \email{a.hajibarat@ph.iut.ac.ir}
   \author{Homayon Anjomshoa}
   \email{h.anjomshoa@ph.iut.ac.ir}

   \affiliation{ Department of Physics, Isfahan University of Technology, Isfahan 84156-83111, Iran}


\date{\today}

\begin{abstract}
	We investigate a class of three parameter metrics that contain both the $\gamma$-metric  and Janis-Newman-Winicour (JNW) metric at special values of the parameters.  To see the effect of the scalar field we derive some properties of this class of metrics such as curvature invariants, the  effective potential, and epicyclic frequencies. We also introduce the five and higher  dimensional forms of the class of metrics in the presence of a scalar field.
	
\begin{description}
\item[PACS numbers]
\end{description}
\end{abstract}

\pacs{Valid PACS appear here}
\maketitle


\section{\label{sec:level1}Introduction}
 Since formulation of general relativity by Albert Einstein in 1915, singular solutions of the theory have attracted attention of a great number of physicists. Based on some famous theorems known as singularity theorems, which were proved by Hawking and Penrose, singularities in GR are inevitable if matter satisfy reasonable energy conditions \cite{Penrose 1}. Because at singularities deterministic equations break down, they are physically problematic. For resolution of this problem Penrose introduced a conjecture called cosmic censorship conjecture (CCC) \cite{Penrose 2}. There are two independent versions of CCC, weak version and strong version. Weak version rules out the possibility of formation of a bare singularity from gravitational collapse of generic non-singular initial data, whereas strong version states that no singularity is visible to any observers. None of these conjectures can be derived from the other one, in addition there is no rigorous mathematical formulation or proof of these conjectures. \\
 
 Singularities in GR are either covered by an event horizon or are naked. Although, based on weak CCC, naked singularities are not allowed, there are  numerous studies that show the outcome of gravitational collapse of physically non-singular initial configuration of matter can be either a black hole or a naked singularity \cite{Joshi,Ori,Christodoulou,Harada,Shapiro}. Due to aforementioned possibility of naked singularities as the end stage of gravitational collapse and their applicability in describing gravitational field  outside of stable configurations of matter, like stars and planets, their study is important and physically relevant. \\
 
  The $\gamma$-metric (also called Zipoy-Voorhees metric or $q$-metric) is an exact solution of the vacuum Einstein's equations with naked singularity that belongs to the class of weyl metrics \cite{Darmois,Erez,Zipoy,Voorhees}. Geometry of this metric was studied in \cite{Herrera,Pastora,Hern,Bonnor}. The interior solution of the $\gamma$-metric was investigated in \cite{Hernandez,Stewart,Magli,Hajibarat}. The structure of circular orbits and epicyclic frequencies of the $\gamma$-metric were studied in \cite{Dadhich}. The higher dimensional 
  extension of the $\gamma$-metric was introduced in \cite{Hajibarat}. Lensing of neutrinos in the weak field limit for the $\gamma$-metric was investigated in \cite{Debasish}. The shadow of the rotating $\gamma$-metric and its comparison with the Kerr metric have been investigated in \cite{Song}. For other studies on the $\gamma$-metric, see also \cite{HerreraPaiva,Richterek,Abdikamalov,Toshmatov,Benavides,Allahyari,Firouzjahi, Chakrabarty,Ozay,Carot,Mustafa,Mangut}.
  
   Another solution with naked singularity is JNW metric, which is a spherically symmetric solution in the presence of a scalar field \cite{Fisher,Janis,Wyman,K.S. Virbhadra}. This metric is a  spherically symmetric solution first introduced in \cite{Fisher,Janis} and later rediscovered in another form in \cite{Wyman}; Then it was shown that these two forms are identical \cite{K.S. Virbhadra}. Some important
   properties of JNW metric was studied in \cite{Chowdhury, Turimov, Makukov}. Gravitational lensing and high energy collisions in this space-time have been investigated in \cite{Virbhadra,Dey, Patil}. The correct form of the rotating JNW and other members of this class of three parameter metrics in the presence of a scalar field can be found in \cite{Mirza}.
  
  Therefore, due to the importance of studying these two specific types of metrics, we derive the most general form of a static metric in the presence of a scalar field. We obtain a class of three parameter exact solutions without spherical symmetry which contain both the $\gamma$-metric (zero scalar field) and JNW metric at certain values of the parameters. We investigate some of the observational signatures of the new class of metrics. It is important that the scalar field can be used as an effective model of dark matter and therefore  this work introduces a significant  class of  metrics which could be useful from an observational perspective. 
 
 This paper is organised as follows: In Sec. II we first obtain a class of three parameter metrics in the presence  of a scalar field. Then in Sec. III, we calculate the effective potential for test particles in this class of space-times. In Sec. IV we study the structure of circular orbits for some special values of the parameters in the class of metrics. Thereafter, in Sec. V the epicyclic frequencies are derived for stable circular orbits. In Sec. VI we introduce the class of three parameter metrics that include both JNW and $\gamma$-metrics  in five dimensions for the first time. Sec. VII is devoted to higher dimensional $(d>5)$ extensions of this class of three parameter metrics.

\section{\label{sec:level2} THE GENERALISED $\gamma$-metric}
The goal of this part is to derive a four dimensional solution of the
  Einstein's equations in the presence of a scalar field. We assume that the scalar field has spherical symmetry but the metric is axially symmetric, which would lead to a class of interesting solutions. The action that we consider is 

 \begin{equation}
	\label{1}
	\mathcal{S}=\int d^4 x ~\sqrt{-g}\left(R- \partial_{\sigma}\varphi(r)\partial^{\sigma}\varphi(r)\right).
\end{equation}

By variation of action in Eq. (\ref{1}), we reach the following equations

\begin{equation}
	\label{metric}
	R_{\alpha\rho}=\partial_{\alpha} \varphi(r) \partial_{\rho} \varphi(r),
\end{equation}
and
 \begin{equation}
	\label{mass1}
	\nabla_{\alpha} \nabla^{\alpha} \varphi(r)=0.
\end{equation}

We assume that the scalar field $\varphi(r)$ depends only on $r$ and  the metric would be in the following form

\begin{equation}
	\label{4A}
	ds^2=-f^\gamma dt^2+f^\mu k^{\nu}(\frac{dr^2}{f}+r^2d\theta^{2})
	+f^{\beta}r^2\sin^2\theta~d\phi^2,
\end{equation}

\noindent where
\begin{eqnarray}
	\label{kye1}
&&f(r)  =1-\frac{2m}{r}, \\
&&k(r,\theta)=1-\frac{2m}{r}+\frac{m^2}{r^2} \mathcal{N}^2(\theta).
\end{eqnarray}

 In this metric $\gamma$, $\mu$, $\nu$, $\beta$ and $\mathcal{N}(\theta)$ are parameters and a function respectively that are going to be determined via solving $R_{\alpha\rho}= \partial_{\alpha} \varphi(r) \partial_{\rho} \varphi(r)$ perturbatively. Therefore we have $R_{tt}=0, R_{\theta\theta}=0$ and $R_{r\theta}=0$. First, we derive $R_{\theta\theta}$ up to the first order in $m$ as follows   
 
 \begin{equation}
 	\label{3}
 	R_{\theta\theta}=-\frac{m (\beta +\gamma -1)}{r}+\mathcal{O}\left(m^2\right),
 \end{equation}

\noindent as a result, we have $\beta=1-\gamma$. 

Now we obtain $R_{r\theta}$  up to the first order of $m$ as below 

\begin{equation}
	\label{4}
	R_{r\theta}=\frac{ (-\beta +\mu +\nu )m \cot \theta}{r^2}+\mathcal{O}\left(m^2\right),
\end{equation} 

\noindent so $\mu+\nu=\beta=1-\gamma$. 

We may use $\beta=1-\gamma$ and $\mu+\nu=1-\gamma$ in the metric in Eq. \eqref{4A} and calculate  $R_{r\theta}$ up to the second order of
 $m$, we obtain

\begin{eqnarray}
	\label{5}
	&&R_{r\theta}=\frac{ (\nu  \mathcal{N}(\theta )) \left(\mathcal{N}'(\theta )-\mathcal{N}(\theta ) \cot \theta \right)m^2}{r^3}+\mathcal{O}\left(m^3\right).~~
\end{eqnarray}

 Then we can  solve $R_{r\theta}=0$ and find $\mathcal{N}(\theta)$ as follows

\begin{equation}
	\label{4}
\mathcal{N}(\theta)=\sin\theta.
\end{equation}

Now we put the achieved parameters and function 
 into metric in \eqref{4A} and derive $R_{rr}$ as follows

\begin{equation}
	\label{11A}
	R_{rr}=\partial_r\varphi(r)\partial_r\varphi(r)=\frac{2 m^2 \left(1-\gamma ^2-\nu\right)}{r^2 (r-2m)^2}.
\end{equation}

By solving Eq. \eqref{11A}, we can obtain the scalar field as follows
\begin{equation}
	\label{7}
	\varphi(r)=\sqrt{\frac{1-\gamma^2-\nu}{2}}~\ln\left(1-\frac{2m}{r}\right).
\end{equation}

It should be noted that at $\nu=1-\gamma^2$ the scalar field vanishes and we have the $\gamma$-metric. Assuming that $\varphi(r)$ to be real, the term under the radical must be positive. It gives us the following constraint on $\mu$ and $\nu$
\begin{eqnarray}
	\label{9}
	\mu\geq\gamma^2-\gamma,~~~ 1-\gamma^2\geq\nu.
\end{eqnarray}

Therefore, we have reached an exact solution of the Einstein's equations with the source of a scalar field. The final form of the metric is

\begin{equation}
	\label{2}
	ds^2=-f^\gamma dt^2+f^\mu k^\nu\big(\frac{dr^2}{f}+r^2d\theta^{2}\big) 
	+f^{1-\gamma}r^2\sin^2\theta~d\phi^2,
\end{equation}

\noindent where
\begin{eqnarray}
	\label{15A}
	&&\mu+\nu=1-\gamma,\\
		&&f(r)  =1-\frac{2m}{r}, \\
		&&k(r,\theta)=1-\frac{2m}{r}+\frac{m^2\sin^2\theta}{r^2}.
\end{eqnarray}  

This solution is interesting because of having three free parameters i.e. mass $m$ and two out of three parameters $\mu$, $\nu$, and $\gamma$. The class of metrics contain both the JNW and $\gamma$-metric: \\

\noindent I. For $\nu=0$, the metric represents the JNW space-time.\\
II. For $\mu=\gamma^2-\gamma$  and $\nu=1-\gamma^2$, we have the $\gamma$-metric and the scalar field vanishes.\\

We can  obtain and study  a lot of new metrics by choosing specific values for $\mu$ and $\nu$ in Eq. \eqref{15A}. In the following sections we are going to investigate the class of metrics and analyse some of its characteristics. It should be noted that in the presence of a scalar field the parameter $\nu$ determines the prolate $ (\nu>0) $ and oblate $ (\nu<0) $ sources.

\section{\label{sec:level2}Basic Characteristics}

The total mass $M=\gamma m$   that corresponds  to the metric in \eqref{2} can be easily obtained by looking at the asymptotic behaviour of the metric components. It is interesting that the naked singularity corresponds to a well-defined total energy that can be observed by an observer at infinity.

\subsection{Singularities}

To determine this metric's singularities we calculate its Ricci scalar and Kretschmann  scalar, because this metric is in the presence of a scalar field its Ricci scalar is non-zero and its singularities can also be read from the Ricci scalar. The Ricci scalar can be obtained in the following form

\begin{equation}
	\label{18A}
	R=\frac{2(\gamma^2+\nu-1)m^2}{r^{2+\gamma-\nu}(r-2m)^{2-\gamma-\nu}\big(r^2-2mr+m^2-m^2\cos^2\theta\big)^\nu}
\end{equation}

From Eq. \eqref{18A} it can be inferred that the metric has singularities at $r=0$ for $\nu-\gamma<2$, $r=2m$ for $\nu + \gamma <2$ and $r=m(1\pm \cos \theta)$ for prolate case with  $\nu>0$. Rate of divergence for the Ricci scalar at $r = 2m$ and $r = 0$ reduces with decreases of $\nu$. For other two singular surface at $r = m(1\pm\cos\theta)$ by increase of $\nu$ rate of divergency  increases. This behavior is correct only for prolate case $(\nu>0)$ where we have singularity at the points. The Kretschmann scalar in its exact form can also be derived; however, it  is too much complicated to be written here. The singularites of both the Kretschmann scalar and the Ricci scalar are exactly at the  same points. For $\nu = 1-\gamma^2$ the Ricci scalar vanishes and we have the $\gamma$-metric. In this special case one has to use the Kretschmann scalar to obtain the singularities.

 It is interesting that the number of singularities and their location are not changed in the presence of a scalar field when compared with the $\gamma$-metric with no scalar field.

\subsection{The Effective Potential of Massive Particles}
The following  Hamiltonian describes the motion of particles in the static space-time metric

\begin{eqnarray}
	\label{11}
	\mathcal{H}=\dfrac{1}{2}g^{\alpha\sigma}p_{\alpha}p_{\sigma}+\dfrac{1}{2}m_{0}^2.
\end{eqnarray}

Here $p_{\alpha}$, the four-momentum of the test particle, is equal to $m_{0}u_{\alpha}$, and $m_{0}$ is the mass of the particle and $u_{\alpha}$ is the four-velocity. Because of the lack of explicit dependence on $t$ and $\phi$ in the Hamiltonian, conservation of the corresponding conjugate momenta $P_t$ and $P_\phi$ will be deduced from Noether's theorem. These conserved conjugate momenta of $t$ and $\phi$ are called energy and angular momentum, respectively,
\begin{eqnarray}
	\label{12}
	 -E&=g_{tt}\dfrac{dt}{d\tau}=p_{t}, \\ 
	 L&=g_{\phi\phi}\dfrac{d\phi}{d\tau}=p_{\phi}.\nonumber
\end{eqnarray}

By using substitution of Eq. \eqref{12} in Eq. \eqref{11} we reach
\begin{eqnarray}
	\label{14}
	\mathcal{H}=\dfrac{1}{2}g^{rr}p_{r}^2+\dfrac{1}{2}g^{\theta\theta}p_{\theta}^2+H_{t\phi}\left(r,\theta\right),
\end{eqnarray}
where
\begin{eqnarray}
	H_{t\phi}\left(r,\theta\right)&=&\dfrac{1}{2}g^{tt}E^2+\dfrac{1}{2}g^{\phi\phi}L^2+\dfrac{1}{2}m_{0}^2, \\ \nonumber
	g^{tt}&=&-f^{-\gamma}, \\ \nonumber
	g^{\phi\phi}&=&\dfrac{f^{\gamma-1}}{r^2\sin^2\theta}.
\end{eqnarray}

By the normalization condition $u_{\alpha}u^{\alpha}=-1$, the following equation can be obtained
\begin{eqnarray}
	\label{20A}
	-\dfrac{2H_{t\phi}\left(r,\theta\right)}{m_{0}^2}=g_{\theta\theta}\left(\dfrac{d\theta}{d\tau}\right)^2+g_{rr}\left(\dfrac{dr}{d\tau}\right)^2.
\end{eqnarray}

If we restrict the motion of the test particles on equatorial plain $\theta=\pi/2$, which is physically justified by our motivation to study the motion of matter on the accretion disc of the assumed astrophysical object described by the metric, the following equation will be achieved
\begin{eqnarray}
	\label{15}
	D(r)\equiv	-\dfrac{2H_{t\phi}\left(r,\theta\right)}{m_{0}^2}=g_{rr}\left(\dfrac{dr}{d\tau}\right)^2,
\end{eqnarray}
where $D(r)$ has the following form

\begin{eqnarray}
	\label{17A}
	D(r)=f^{-\gamma}\left(E^2-V_{eff}(r)\right).
\end{eqnarray}

\noindent and the effective potential will be read as follows
\begin{eqnarray}
	\label{24}
	V_{eff}(r)=\dfrac{L^2}{r^2}f^{2\gamma-1}+f^{\gamma}
\end{eqnarray}

\begin{figure}[h]
	\centering
	\includegraphics[width=0.4\textwidth]{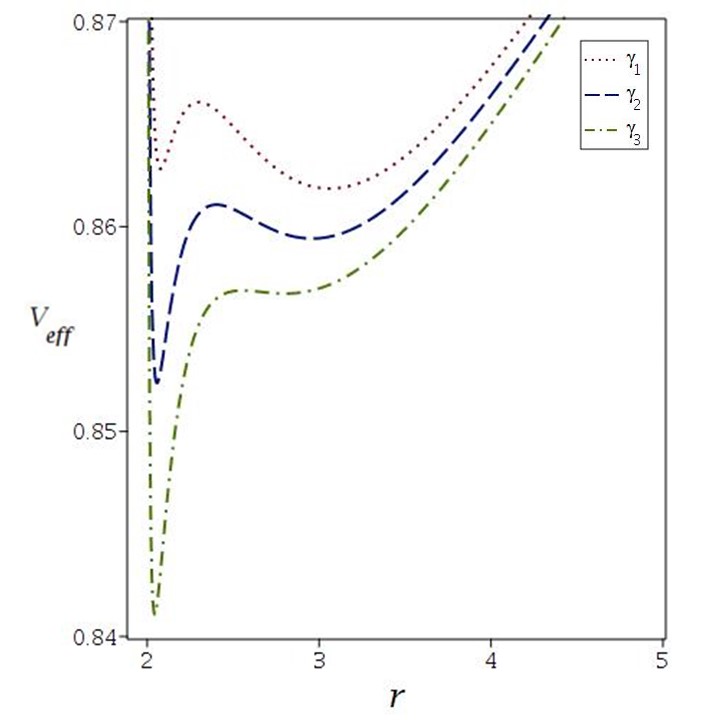}
	\caption[]{\it The effective potential in Eq. \eqref{24} for $L=1.66$, $m=1$ and $\gamma\in\left(\frac{1}{\sqrt{5}},\frac{1}{2}\right)$. $\gamma_1=0.460$, for the dotted line, $\gamma_2=0.462$ for the dashed line and  the dash-dotted line for $\gamma_3=0.464$.}
	\label{fig:Veff1}
\end{figure}

\begin{figure}[h]
	\centering
	\includegraphics[width=0.5\textwidth]{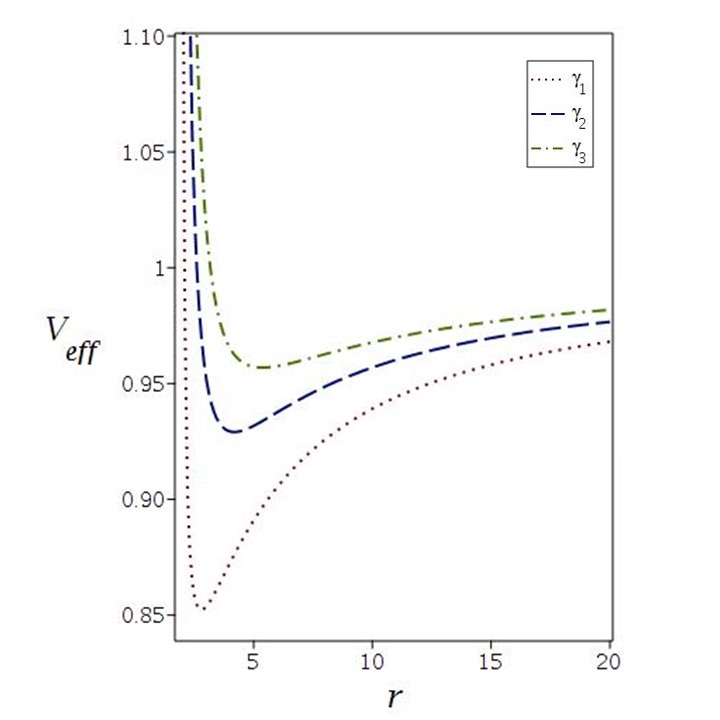}
	\caption[]{\it The effective potential of \eqref{24} for $L=1$, $m=1$ and $\gamma\notin\left(\frac{1}{\sqrt{5}},\frac{1}{2}\right)$. $\gamma_1=1/3$ for the dotted line, $\gamma_2=1/4$ for the dashed line and  the dash-dotted line for $\gamma_3=1/5$. }
	\label{fig:Veff}
\end{figure}

It is an interesting observation that the effective potential depends only on $\gamma$. The effective potential for a massive particle has two typical behaviors for different values of $\gamma$. The effective potentials are depicted in Fig. \ref{fig:Veff1} and Fig. \ref{fig:Veff} for different values of $\gamma$ and $L$.

From Fig. \ref{fig:Veff1} it can be seen that the effective potential has two minimum for $\gamma\in\left(1/\sqrt{5},1/2\right)$, and from Fig. \ref{fig:Veff} it can be seen that effective potential has only one minimum for $\gamma\notin\left(1/\sqrt{5},1/2\right)$.
\subsection{The Effective Potential for Massless Particles}
Now we are going to calculate the effective potential $\mathcal{V}_{eff}$ for massless 
 particles. The metric components are not dependent on
 $t$ and $\phi$ and we have two killing vectors $\mathcal{K}_{1}^{\alpha}$ and $\mathcal{K}_{2}^{\alpha}$ as follows 
\begin{eqnarray}
	\label{33}
	\mathcal{K}_{1}^{\alpha}&=&\left(1,0,0,0\right), \\ \nonumber
	\mathcal{K}_{2}^{\alpha}&=&\left(0,0,0,1\right).
\end{eqnarray}

Therefore corresponding to the two Killing vectors, we have two conserved quantities called $E$ and $L$ that can be written as follows
\begin{eqnarray}
	\label{17}
	-E&=&g_{tt}\dfrac{dt}{d\tau}=\mathcal{K}_{1\alpha}\frac{dx^{\alpha}}{d\tau},\\ \nonumber
	L&=&g_{\phi\phi}\dfrac{d\phi}{d\tau}=\mathcal{K}_{2\alpha}\frac{dx^{\alpha}}{d\tau}.\\ \nonumber
\end{eqnarray}

In addition, by using $u_{\alpha}u^{\alpha}=0$ for null geodesic and restrict ourselves to the equatorial plane $\theta=\pi/2$, we have
\begin{eqnarray}
	\label{18}
	g_{tt}\left(\dfrac{dt}{d\tau}\right)^2+g_{rr}\left(\dfrac{dr}{d\tau}\right)^2+g_{\phi\phi}\left(\dfrac{d\phi}{d\tau}\right)^2=0.
\end{eqnarray}

Using Eq. \eqref{17} and after simplification, we obtain
\begin{eqnarray}
	\label{19}
	\dfrac{E^2}{g_{tt}}+\dfrac{L^2}{g_{\phi\phi}}+g_{rr}\left(\dfrac{dr}{d\tau}\right)^2=0.
\end{eqnarray}

Therefore the effective potential for massless particles can be read as 
\begin{eqnarray}
	\label{20}
	\mathcal{V}_{eff}=-\dfrac{1}{2}\dfrac{g_{tt}}{g_{\phi\phi}}L^2=\dfrac{L^2}{r^2}f^{2\gamma-1}.
\end{eqnarray}

\begin{figure}[h]
	\centering
	\includegraphics[width=0.4\textwidth]{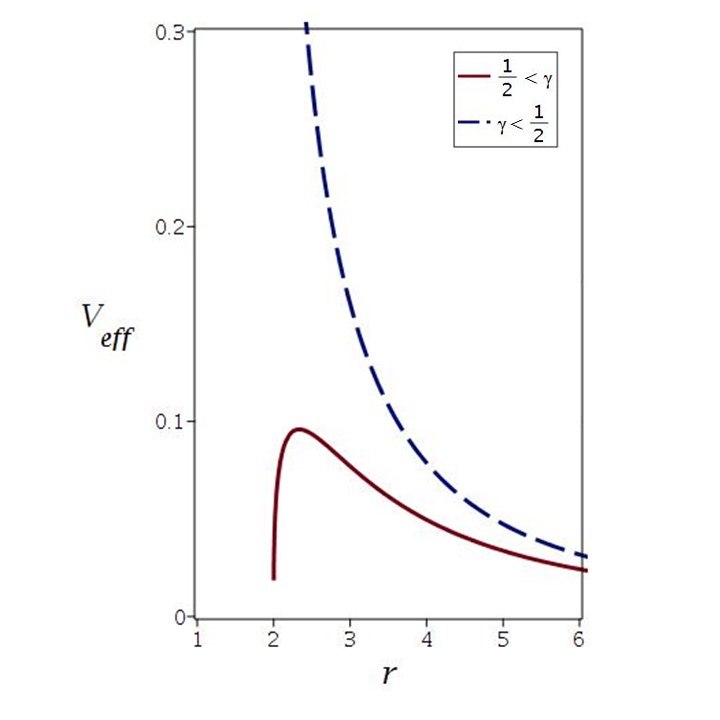}
	\caption[]{\it The effective potential of Eq. \eqref{20} for $L=1$, $m=1$. The effective potential diverges to $-\infty$ at $r=2m$ and two values of $\gamma$, the red continuous line  for $\gamma>1/2$ and the dashed line for $\gamma<1/2$. The effective potential diverges to $+\infty$ at $r=2m$.}
	\label{fig:figure 1}
\end{figure}
As it can be seen in Fig. \ref{fig:figure 1} for $\gamma<1/2$, the effective potential of a massless particles has no minimum or maximum, but for $\gamma>1/2$ it has one maximum. In the following section, we investigate the physical implication of different  behaviors  of the effective potential for several values of $\gamma$.
\section{\label{sec:level3}Circular Orbits }
In this section, we investigate the radii of circular orbits in the equatorial plane as they explain the behavior of gas particles in accretion disks around the massive 
 objects. It should be noted that in this kind of circular motion as both $r$ and $\theta$ are constant we don't see the effects of the scalar field. This section is  just a review \cite{Dadhich}  and will be used  in the next section that  the effect of the scalar field can be seen in the epicyclic frequencies. Radii of orbits for specific values of $E$ and $L$ can be determined by solving the following equations
\begin{eqnarray}
	\label{25}
	V_{eff}&=&E^2, \\ \nonumber
	\dfrac{d}{dr}\left(V_{eff}\right)&=&0.
\end{eqnarray}

For a circular orbit at $r$, we must choose $E$ and $L$ in such a way that  Eq. \eqref{25} will be satisfied and have a solution for $r$. By putting Eq. \eqref{24} in \eqref{25} we can find $E$ and $L$ as follows

\begin{eqnarray}
	\label{26}
	L=\pm r\left(\dfrac{r}{m\gamma}-\dfrac{1}{\gamma}-2\right)^{-\frac{1}{2}}f^{\frac{1-\gamma}{2}},
\end{eqnarray}
\begin{eqnarray}
	\label{27A}
	E=\left(\dfrac{r}{m\gamma}-\dfrac{1}{\gamma}-2\right)^{-\frac{1}{2}}\left(\dfrac{r}{m\gamma}-\dfrac{1}{\gamma}-1\right)^{\frac{1}{2}}f^{\frac{\gamma}{2}}.
\end{eqnarray}

From \eqref{26} and \eqref{27A} it can be seen that the regions of our metric which allow circulars orbits are given by
\begin{eqnarray}
	\label{27}
	~~~~\textnormal{if} \quad   \gamma\geq\dfrac{1}{2}\quad \Rightarrow \quad r>m\left(2\gamma+1\right),
\end{eqnarray} 
and
\begin{eqnarray}
	\label{28}
	\textnormal{if} \quad   0<\gamma<\dfrac{1}{2}\quad \Rightarrow \quad r>2m.
\end{eqnarray}

Not all circular orbits are stable. Stability of a circular orbit is going to be determined by the positivity of $\dfrac{d^2}{dr^2}\left(V_{eff}\right)$. By calculating $\dfrac{d^2}{dr^2}\left(V_{eff}\right)$ and using Eq. \eqref{26}, we have
\begin{eqnarray}
	\label{29}
		\dfrac{d^2}{dr^2}\left(V_{eff}\right)&=&F(r,\gamma)\dfrac{-2\gamma m}{r^2}\\ \nonumber
		&+&F(r,\gamma)\dfrac{12\gamma m^2\left(\gamma+\dfrac{1}{3}\right)}{r^3}\\ \nonumber
		&+&F(r,\gamma)\dfrac{-8\gamma m^3\left(\gamma^2+\dfrac{3}{2}\gamma+\dfrac{1}{2}\right)}{r^4},
\end{eqnarray}
where
\begin{eqnarray}
	\label{30}
	F(r,\gamma)=\dfrac{f^{\gamma-2}}{2m\gamma+m-r}.
\end{eqnarray}

Eq. \eqref{29} has two roots at $r_{isco,+}$ and $r_{isco,-}$, which are given by
\begin{eqnarray}
	\label{31}
	r_{isco,\pm}=m\left(1+3\gamma\pm\sqrt{5\gamma^2-1}\right).
\end{eqnarray}

$\dfrac{d^2}{dr^2}\left(V_{eff}\right)$ is positive for $r\notin\left[r_{isco,-},r_{isco,+}\right]$ and negative or zero otherwise. Therefore, circular orbits for $r\in\left[r_{isco,-},r_{isco,+}\right]$ and $r\notin\left[r_{isco,-},r_{isco,+}\right]$ are unstable and stable, respectively.

Unstable circular orbits are called bound or unbound if their small deviation leads to bound or unbound motion respectively. The boundness condition is $E<1$ (because in this case test particle cannot escape to infinity). So we define $r_{mb}$ as a marginally bound orbit whose outward departures lead to unbound motion and inward ones to bound motion. Therefore, using $E=1$ in Eq. \eqref{27A}  we have the following condition for $r_{mb}$
\begin{eqnarray}
	\label{32}
	r_{mb}-m\left(1+2\gamma\right)-
	\left(r_{mb}-m\left(1+\gamma\right)\right)f\left(r_{mb}\right)^\gamma=0.~~~~~
\end{eqnarray}
\begin{figure}[h]
	\centering
	\includegraphics[width=0.45\textwidth]{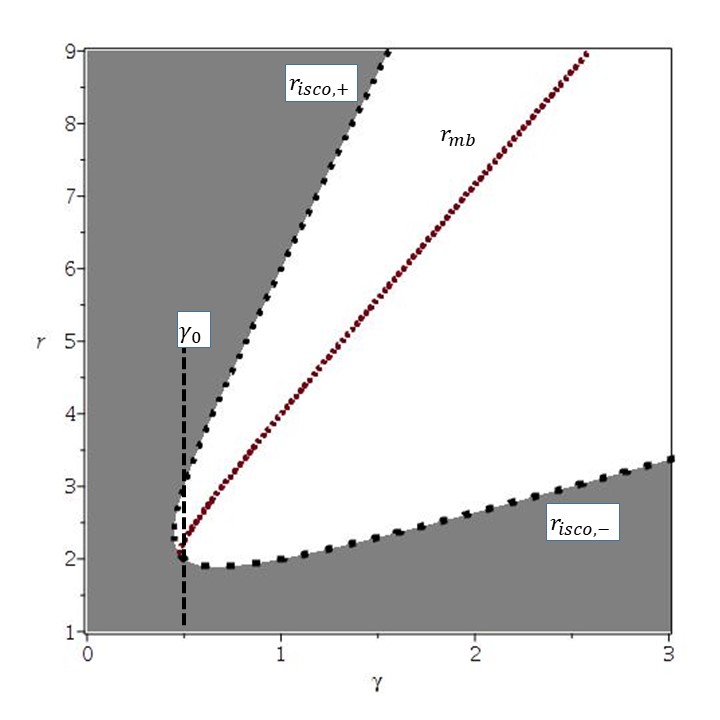}
	\caption[]{\it $r_{isco,\pm}$ and $r_{mb}$ in Eq. \eqref{31} and Eq. \eqref{32} respectively for $0<L<0.6$ and $m=1$.For $r>2m$ the grey area represents  stable circular orbits. The white are between $r_{isco,+}$ and $r_{isco,-}$ is for unstable circular orbits and $r_{mb}$ is for marginally bound circular orbits, for $r<r_{mb}$ small deviation of the orbits leads to bound non-circular motion.}
	\label{fig:RMB}
\end{figure}
We can see from Fig.\ref{fig:RMB} that for $\gamma<1/\sqrt{5}$ all circular orbits are stable, for $1/\sqrt{5}<\gamma<\gamma_{0}$ unstable orbits are unbounded and for $\gamma_{0}<\gamma$ unstable orbits can be  bound. 

Until now, we have studied circular orbits for massive test particles. For answering whether there are circular orbits for massless particles we must use $\mathcal{V}_{eff}$ in Eq. \eqref{20}.

In contrast to the massive case, one notices after putting Eq. \eqref{20} in Eq. \eqref{25} that only $E/L$ can be determined and exact values of $E$ and $L$  remain unknown. From $\dfrac{d}{dr}\left(\mathcal{V}_{eff}\right)=0$, we can determine $r_{ps}$ which is the circular orbit for a massless particle
\begin{eqnarray}
	\label{34}
	r_{ps}=m\left(2\gamma+1\right).
\end{eqnarray}

We have circular orbits for massless particles when $\gamma>1/2$. Similar to massive particles by calculating $\dfrac{d^2}{dr^2}\left(\mathcal{V}_{eff}\right)|_{r=r_{ps}}=0$, we see that $r_{ps}$ is an unstable circular orbit. 

\begin{figure}[h]
	\centering
	\includegraphics[width=0.4\textwidth]{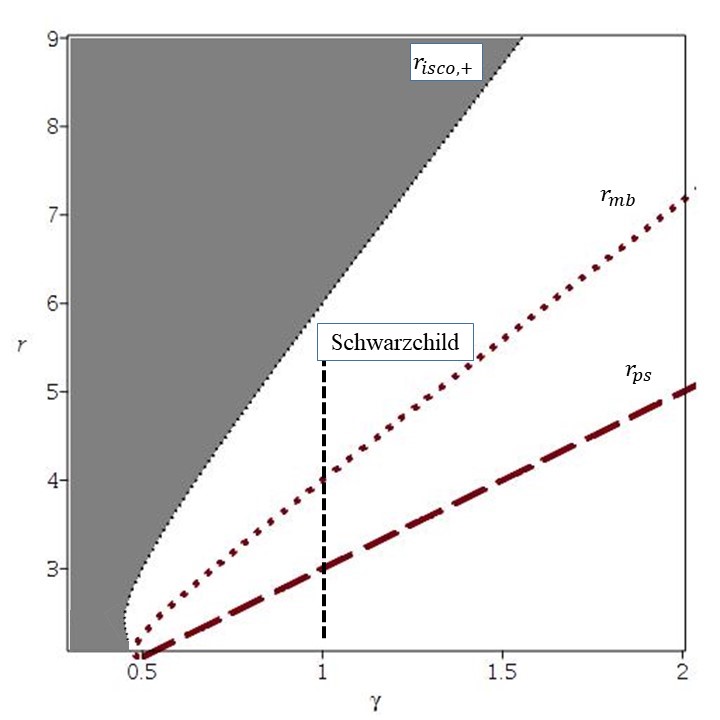}
	\caption[]{\it $r_{isco,+}$, $r_{mb}$  and $r_{ps}$ in Eq. \eqref{31}, Eq. \eqref{32} and Eq. \eqref{34} respectively for $L=1$ and $m=1$. For $r<r_{ps}$ there is no circular orbits. $r_{ps}$ is where we have null circular orbits for massless particles.}
	\label{fig:RPS}
\end{figure}

The gray area in Fig.\ref{fig:RPS} shows stable circular orbits and the area between $r_{isco,\pm}$ and $r_{ps}$ is for unstable circular orbits. Moreover $r_{ps}<r<r_{mb}$ is for bound unstable circular orbits.
\section{\label{sec:level2}Epicyclic Frequencies}

 Due to the astrophysical importance of epicyclic frequencies, our aim in this section is to calculate epicyclic frequencies and study the effect of the scalar field
  in these frequencies.

For simplicity, we consider epicyclic oscillations around circular orbits for  purely vertical  $\theta=\theta_{0}+\delta\theta$, $\delta r=0$ and purely radial $r=r_{0}+\delta r$, $ \delta\theta=0$ oscillations. By applying these conditions to equation \eqref{20A} we reach to 

\begin{eqnarray}
	\label{43A}
	-\frac{2}{g_{rr}}\frac{\partial^2H_{t\phi}}{\partial r^2}\delta r^2=\big(\frac{d\delta r}{d\tau}\big)^2, \\
	\label{43B}
	-\frac{2}{g_{\theta\theta}}\frac{\partial^2H_{t\phi}}{\partial \theta^2}\delta\theta^2=\big(\frac{d\delta \theta}{d\tau}\big)^2,
\end{eqnarray}

\noindent which result into a periodic solution with frequencies 

\begin{eqnarray}
	\label{40}
	\Omega_{r}^{2}=\dfrac{\dfrac{\partial^{2}}{\partial r^{2}}\big(H_{t\phi}\left(r,\theta\right)\big)\vert_{r_{0}}}{g_{rr}},
\end{eqnarray}
and
\begin{eqnarray}
	\label{41}
	\Omega_{\theta}^{2}=\dfrac{\dfrac{\partial^{2}}{\partial \theta^{2}}\big(H_{t\phi}\left(r,\theta\right)\big)\vert_{\theta_{0}}}{g_{\theta\theta}}.
\end{eqnarray}

Here $r_{0}$ is the radius of a circular orbit. From our linearized equations, we can see that solutions are unbounded when frequencies \eqref{40} and \eqref{41} are imaginary, which takes place at unstable circular orbits. In our calculations, for the sake of simplicity, we only consider circular orbits at latitudinal plane ($\theta_{0}=\pi/2$).

In addition to these frequencies, there are another important ones which is the frequencies of the circular motion of test particles, commonly called Keplerian frequency. Keplerian frequency can be read from  \eqref{17} as follows
\begin{eqnarray}
	\label{42A}
	\Omega_{\phi}^{2}=\dot{\phi}^{2}=\left(g^{\phi\phi}L\right)^2.
\end{eqnarray}

Because these frequencies are being measured with respect to the comoving observer, we must take into account the redshift effect. To obtain what an observer at infinity measures, we must divide  $\Omega_r$, $\Omega_\theta$ and $\Omega_\phi$ by square of red shift factor $u^0$ and reach to $\omega_r$, $\omega_\theta$ and $\omega_\phi$ which are epicyclic frequencies as observed at infinity.  Square redshift factor for our metric is as follows 
\begin{eqnarray}
	\label{42}
	(u^{0})^{2}=\dfrac{r-m-\gamma m}{r-m-2\gamma m}f^{-\gamma}.
\end{eqnarray}

After substitution of $E$ and $L$ (Eq. \eqref{26} and Eq. \eqref{27A}) in Eqs. \eqref{40},  \eqref{41} and \eqref{42A}, we obtain explicit forms of  epicyclic frequencies. The behavior of these frequencies can be abbreviated in the following way
\begin{eqnarray}
	\label{43}
	&&\omega_{r}^{2}=\left(\dfrac{m-r}{r}\right)^{2\left(\gamma+\mu-1\right)}\\ \nonumber &&\times
	\dfrac{m\gamma\left[\left(4\gamma^{2}+6\gamma+2\right)m^{2}-2mr\left(3\gamma+1\right)+r^{2}\right]}{r^{4}\left(r-m-m\gamma\right)}f^{\gamma-\mu-1},
\end{eqnarray}

\begin{figure}
	\centering
	\includegraphics[width=0.5\textwidth]{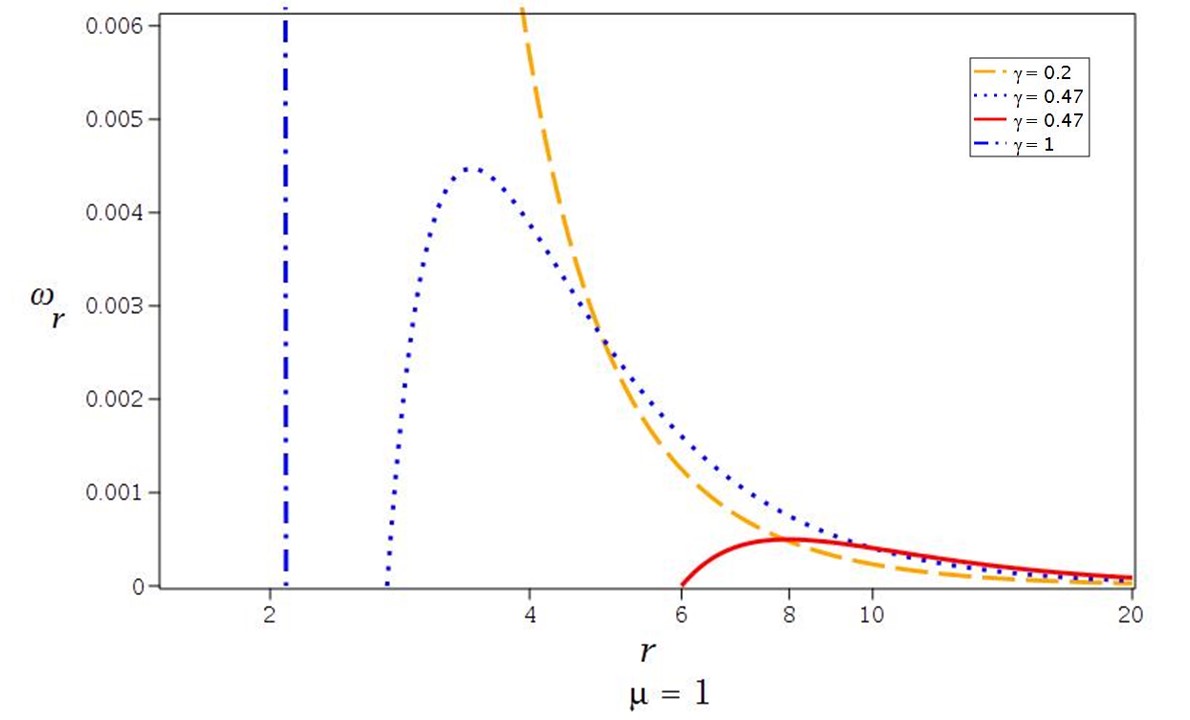}
	\caption[]{\it Typical behaviors of $\omega_r$ for different values of $\gamma$ and $\mu$ based on Eq. \eqref{43} with $m=1$. Based on  Eq. \eqref{15A} and our values for $\mu$ and $\gamma$ all of these cases are for  a case with $\nu<0$ which is an oblate one, but as we have mentioned in the text, behavior similar to these can be seen for prolate cases.}
	\label{fig:figure 11}
\end{figure}

In the $\gamma$-metric, $\gamma$ can be considered as a deformation parameter that measures deviation from spherical case; for $\gamma\neq0$ there is no spherical symmetry. For the case of $\gamma$-metric $\gamma>1$ and $\gamma<1$ correspond to oblate and prolate sources respectively. However, in the presence of a scalar field the parameter that measures deviation from spherical symmetry is $\nu$, where $\nu\neq1$ shows lack of spherical symmetry. In the presence of a scalar field $\nu<0$ and $\nu>0$ determine oblate and prolate sources, respectively.\\
For $\gamma<1/\sqrt{5}$ and $\gamma-\mu-1<0$, $\omega_r$ diverges at $r=2m$, these behaviors takes place for both oblate and prolate cases. When $1/\sqrt{5}<\gamma<1/2$, $\omega_r$ is not defined for $r_{isco,-}<r<r_{isco,+}$ and if $\gamma-\mu-1<0$ it goes as before to infinity at $r=2m$, again this behavior is the same for both oblate and prolate cases. For $\gamma>1/2$, $\omega_r=0$ at $r=r_{isco,+}$, for both oblate and prolate cases. Typical behaviors of $ \omega_r$  are  shown in Fig. \eqref{fig:figure 11}. The other frequency $ \omega_\theta$   can be obtained as follows

\begin{eqnarray}
	\label{44}
	\omega_{\theta}^{2}=\left(\dfrac{m-r}{r}\right)^{2\left(\gamma+\mu-1\right)}\times 
	\dfrac{m\gamma}{r^{2}\left(r-m-m\gamma\right)}f^{\gamma-\mu},~~~
\end{eqnarray}

\begin{figure}
	\centering
	\includegraphics[width=0.5\textwidth]{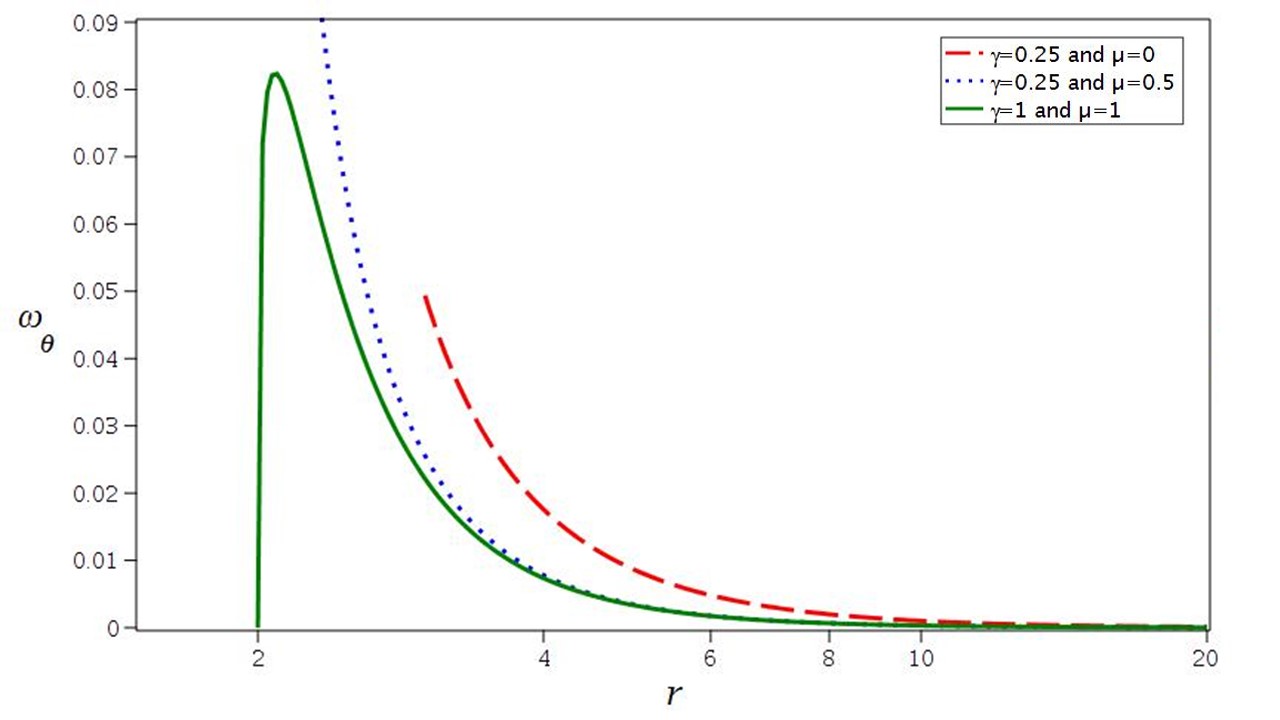}
	\caption[]{\it Typical behaviors of $\omega_\theta$ for different values of $\gamma$ and $\mu$ based on Eq. \eqref{44} with $m=1$. Dashed and dotted lines are for prolate cases with positive $\nu$ and continuous red line is for an oblate case with negative $\nu$. Again we emphasize that all of these typical behaviors can be observed for both oblate and prolate cases.}
	\label{fig:figure 2}
\end{figure}

If $\gamma<1/\sqrt{5}$ and $\gamma<\mu$, $\omega_\theta$ goes to infinity at $r=2m$, this behavior is only for prolate case. for $\mu<\gamma$ we have $\omega_\theta=0$ at $r=2m$, this behavior takes place in both oblate and prolate cases. For $1/\sqrt{5}<\gamma<1/2$, $\omega_\theta$ is not defined for $r_{isco,-}<r<r_{isco,+}$ and for $\lambda<\mu$, $\omega_{\theta}$ goes to infinity, this case happens only for prolate case. For $1/\sqrt{5}<\gamma<1/2$ and $\gamma>\mu$, $\omega_{\theta}$ is not defined for $r_{isco,-}<r<r_{isco,+}$ and $\omega_{\theta}=0$ at $r=2m$, this behavior can be seen in both oblate and prolate cases. For $\gamma>1/2$, $\omega_\theta$ is only defined for $r>r_{isco,+}$ and it is a constant at $r_{isco,+}$, this behavior is for both oblate and prolate cases. These typical behaviors are shown in Fig. \eqref{fig:figure 2}.

\begin{eqnarray}
	\label{45}
	\omega_{\phi}^{2}=
	\dfrac{\left[m\gamma\left(r-m-2m\gamma\right)\right]^{1/2}}{r\left(r-m-m\gamma\right)}f^{\frac{1}{2}\left(3\gamma-1\right)}.
\end{eqnarray}

\begin{figure}[h]
	\centering
	\includegraphics[width=0.4\textwidth]{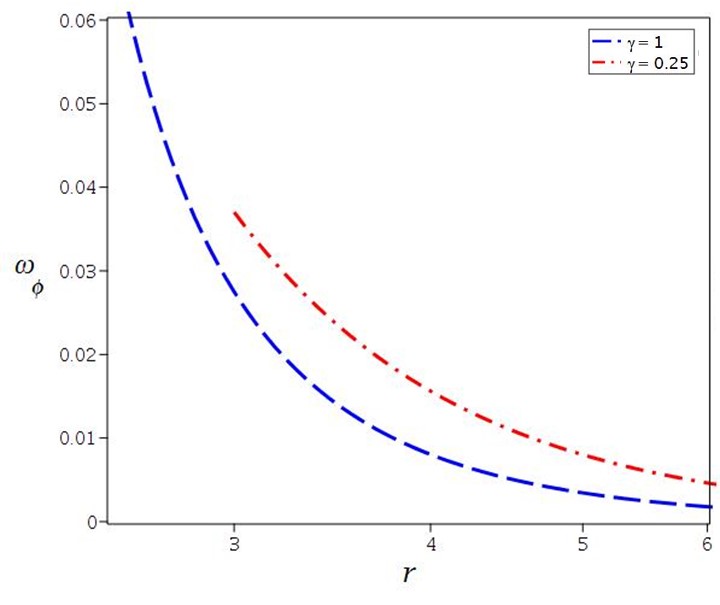}
	\caption[]{\it Typical behaviors of $\omega_\phi$ for different values of $\gamma$ and $\mu$ based on Eq. \eqref{45} with $m=1$. Because of Eq. \eqref{45} being independent of $\mu$ both of these cases can represent an oblate or a prolate case.}
	\label{fig:figure 3}
\end{figure}

For both oblate and prolate cases the following behaviors can be observed for $\omega_{\phi}$. For $\gamma<1/3$, $\omega_\phi$ diverges at $r=2m$. For $1/3<\gamma<1/\sqrt{5}$, $\omega_\phi$ vanish at $r=2m$. For $1/\sqrt{5}<\gamma<1/2$, $\omega_\phi$ is not defined for $r_{isco,-}<r<r_{isco,+}$ and $\omega_\phi$ vanishes at $r=2m$. For $\gamma>1/2$,   $\omega_\phi$ is defined only for $r>r_{isco,+}$ and at $r_{isco}$ it is a non-zero constant. Similar to oblate case for $\omega_\theta$, $\omega_\phi$ in oblate case is only defined for $r>r_{isco,+}$ and goes to a non-zero constant with no maximum. These typical behaviours are shown in Fig. \eqref{fig:figure 3}. Therefore observation of the epicyclic frequencies is important and could be used as a method to see the effects of the scalar field or dark matter.

\section{\label{sec:level2}five DIMENSIONAL METRICS
}

We use the same method as in Sec. II to derive the five-dimensional extensions of the three parameter metric in the presence of a scalar field. First, we assume the following form for the five-dimensional metric

\begin{eqnarray}
	\label{54}
	ds^2=&-&f^\gamma dt^2+f^\mu k^\nu\big(\frac{dr^2}{f}+r^2d\theta^{2}\big) 	\nonumber\\&+&r^2f^\beta\big(\sin^2\theta d\phi^2+\cos^2\theta d\psi^2\big),
\end{eqnarray}

\noindent with

\begin{eqnarray}
	\label{55}
		&&f(r)  =1-\frac{2m}{r^{2}}, \\
		&&k(r,\theta)=1-\frac{2m}{r^2}+\frac{m^2}{r^4} n^2(\theta).
\end{eqnarray}  

\noindent where $\mu$, $\nu$ and $\beta$ are unknown parameters and $n(\theta)$ is an unknown function which, as before, should be determined by solving the equations of $R_{\alpha\xi}= \partial_{\alpha} \Phi(r) \partial_{\xi} \Phi(r)$. We derive  $R_{r\theta}$ and $R_{\theta\theta}$ up to the first order in $m$ as follows

\begin{eqnarray}
	\label{56A}
	&&R_{r\theta}=\frac{4(-\beta+\mu+\nu)\cot(2\theta) m}{r^3}+\mathcal{O}(m^2),\\
	\label{57A}
	&&R_{\theta\theta}=\frac{2(1-2\beta-\gamma)m}{r^2}+\mathcal{O}(m^2).
	\end{eqnarray}

By setting $R_{r\theta}=0$ and $R_{\theta\theta}=0$, we obtain  

\begin{eqnarray}
	\label{56B}
	&&\mu+\nu=\beta,\\
	\label{57B}
	&&\beta=\frac{1-\gamma}{2}.
\end{eqnarray}

Now by using Eqs. \eqref{56B} and \eqref{57B} in the metric and then calculating $R_{r\theta}$  up to the second order in $m$, we have

\begin{equation}
	\label{58A}
	R_{r\theta}=\frac{2\nu n(\theta)\big(-2\cot(2\theta)n(\theta)+\frac{dn(\theta)}{d\theta}\big)m^2}{r^5}.
\end{equation}

Due to the fact that the scalar field $\Phi(r)$ depends only on $r$, we have $R_{r\theta}=0$, and therefore $n(\theta)$ can be derived as follows  

\begin{eqnarray}
		&&n(\theta)=\sin(2\theta).
\end{eqnarray}

The final metric form is as follows

\begin{eqnarray}
	\label{5555}
	ds^2&=&-f^\gamma dt^2+f^\mu k^\nu\big(\frac{dr^2}{f}+r^2d\theta^{2}\big) 	\nonumber\\&+&r^2f^\beta\big(\sin^2\theta ~d\phi^2+\cos^2\theta~ d\psi^2\big),
\end{eqnarray}

\noindent where ($\mu+\nu=\beta$ and $\beta=\frac{1-\gamma}{2}$),

\begin{eqnarray}
	\label{5555555}
	&&f(r)=1-\frac{2m}{r^{2}}, \\
	&&k(r,\theta)=1-\frac{2m}{r^2}+\frac{m^2\sin^2(2\theta)}{r^4} .
\end{eqnarray}  

This solution satisfies the Einstein's equation in the presence
 of a scalar field in the following form
 
 \begin{equation}
 	\label{55A}
 	R_{rr}=\partial_r\Phi(r)\partial_r\Phi(r)=\frac{2m^2(3-3\gamma^2-4\nu)}{r^2(r^2-2m)^2}.
 \end{equation}

Therefore we can solve Eq. \eqref{55A} and derive the scalar field as follows
\begin{eqnarray}
	\label{56}
	\Phi(r)&=&\sqrt{\frac{\frac{3}{4}(1-\gamma^2)-\nu}{2}}~\ln\left(1-\frac{2m}{r^{2}}\right),\nonumber\\
	&=&\sqrt{\frac{\mu-\frac{1}{4}(3\gamma^2-2\gamma-1)}{2}}~\ln\left(1-\frac{2m}{r^{2}}\right).
\end{eqnarray}

The scalar field $\Phi(r)$ also satisfies the following equation
\begin{equation}
	\label{57}
	\nabla_{\alpha} \nabla^\alpha \Phi(r)=0.
\end{equation}

In order for the scalar field to be real, the following restrictions must be satisfied with $\mu$ and $\nu$

\begin{eqnarray}
	\label{65A}
	&&\mu\geq\frac{1}{4}(3\gamma^2-2\gamma-1),\\
	&&\nu\leq\frac{3}{4}(1-\gamma^2).
\end{eqnarray}

Like the four-dimensional solution, the five-dimensional case also includes both the $\gamma$-metric and the JNW metric as follows:\\

\noindent I. For $\nu=0$, the metric becomes the five-dimensional JNW metric \cite{Abdolrahimi}.\\
II. For $\mu=\frac{1}{4}(3\gamma^2-2\gamma-1)$  and $\nu=\frac{3}{4}(1-\gamma^2)$, we obtain the five-dimensional $\gamma$-metric and the scalar field vanishes \cite{Hajibarat}.\\

We may derive exact forms of both the Ricci and Kretschmann scalars. As before the Kretschmann scalar's exact form is too long and we avoid writing its exact form; however, the singularities of both the Ricci and Kretschmann scalars are the same. The Ricci scalar can be derived in the following form

\begin{equation}
	R=\frac{2m^2(2\gamma+4\mu+1-3\gamma^2)}{r^{\gamma-2\nu+3}(r^2-2m)^{\frac{3-\gamma-2\nu}{2}}\big(r^4-2mr^2+m^2\sin^2(2\theta)\big)^\nu},
\end{equation}
\\
\noindent
which has singularities at $r=0$ for $2\nu-\gamma<3$, $r=\sqrt{2m}$ for $\gamma+2\nu<3$ and $r=\sqrt{m(1\pm\cos(2\theta))}$ for $\nu>0$. The Ricci scalar vanishes at $\mu=\frac{3\gamma^2-2\gamma-1}{4}$ where we have the five dimentional $\gamma$-metric and its singularities can be obtained by calculating the Kretschmann scalar \cite{Hajibarat}.

\subsection{The effective potential}

The effective potential contains a rich amount of useful information. So in this section, we calculate it and investigate its behavior. For metric in \eqref{5555}, we have three Killing vector

\begin{eqnarray}
	\label{59}
	K^\alpha&=&\big(\partial_t\big)^\alpha=(1,0,0,0,0),\\
	\label{60}
	R^\alpha&=&\big(\partial_\phi\big)^\alpha=(0,0,0,1,0),\\
	\label{61}
	S^\alpha&=&\big(\partial_\psi\big)^\alpha=(0,0,0,0,1).
\end{eqnarray}

By lowering the upper indices, we have
\begin{eqnarray}
	\label{62}
	K_\alpha&=&g_{\alpha\rho}K^\rho=(-f^\gamma,0,0,0,0),\\
	\label{63}
	S_\alpha&=&g_{\alpha\rho}S^\rho=(0,0,0,f^{\frac{1-\gamma}{2}}r^2\sin^2\theta,0),\\
	\label{64}
	R_\alpha&=&g_{\alpha\rho}R^\rho=(0,0,0,0,f^{\frac{1-\gamma}{2}}r^2\cos^2\theta).
\end{eqnarray}

Also we have
\begin{eqnarray}
	\label{65}
	E&=&-K_\alpha\frac{dx^\alpha}{d\lambda}=f^\gamma\big(\frac{dt}{d\lambda}\big),\\
	\label{66}
	L_\phi&=&R_\alpha\frac{dx^\alpha}{d\lambda}=r^2f^{\frac{1-\gamma}{2}}\sin^2\theta\big(\frac{d\phi}{d\lambda}\big),\\
	\label{67}
	L_\psi&=&S_\alpha\frac{dx^\alpha}{d\lambda}=r^2f^{\frac{1-\gamma}{2}}\cos^2\theta\big(\frac{d\psi}{d\lambda}\big).
\end{eqnarray}

By using the normalized condition $\epsilon=-g_{\alpha\rho}\dot{x^\alpha}\dot{x^\rho}$ and also using \eqref{65}, \eqref{66} and \eqref{67}, we obtain (for the case of $\theta$ not being fixed)

\begin{equation}
	\label{68}
	\frac{1}{2}\Big(\frac{dr}{d\lambda}\Big)^2+V_{eff}(r,\theta)=\frac{E^2}{2}\frac{f^{1-\mu-\gamma}}{k^\nu},
\end{equation}

\noindent where the effective potential is

\begin{eqnarray}
	\label{69}
	V_{eff}(r,\theta)&=&\frac{\epsilon}{2}\frac{f^{1-\mu}}{k^\nu}+r^2\frac{f}{2}\dot{\theta}\nonumber\\&+&\frac{1}{2r^2}\frac{f^{1+\gamma-2\mu}}{k^\nu}\Big(\frac{L_\phi^2}{\sin^2\theta}+\frac{L_\psi^2}{\cos^2\theta}\Big).
\end{eqnarray}

For $\theta=\pi/2$ (i.e. $L_\psi=0$), we have

\begin{equation}
		\label{70}
	\frac{1}{2}\Big(\frac{dr}{d\lambda}\Big)^2+V_{eff}(r)=\frac{E^2}{2}f^{\frac{1-\gamma}{2}},
\end{equation}

\noindent where 

\begin{equation}
		\label{71}
	V_{eff}(r)=\frac{\epsilon}{2}f^{\frac{\gamma+1}{2}}+\frac{L_\phi^2}{2r^2}f^{\frac{1-2\mu+3\gamma}{2}}.
\end{equation}

\begin{figure}[h]
	\centering
	\includegraphics[width=0.4\textwidth]{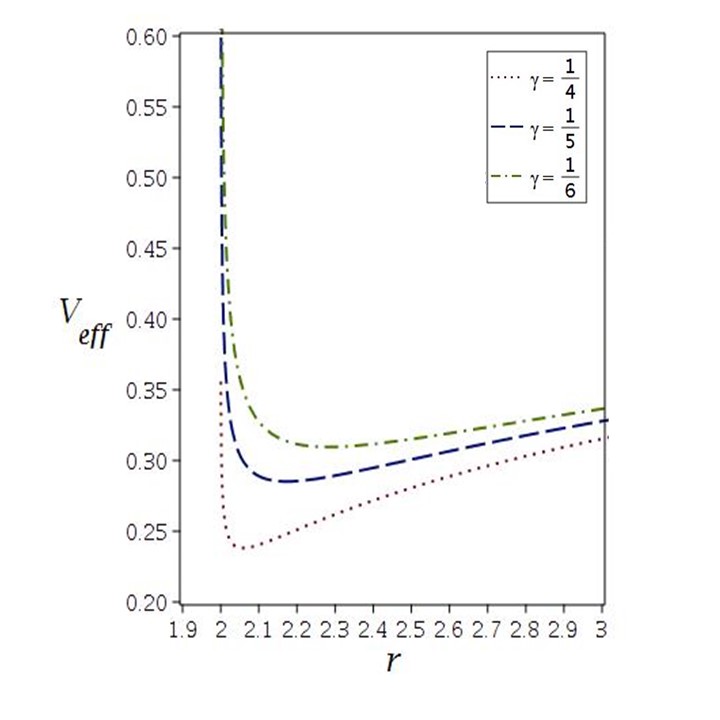}
	\caption[]{\it Effective potential of Eq. \eqref{71} for $\gamma=1/4, 1/5$ and $1/6$ and $\mu=1$.}
	\label{fig:figure 4}
\end{figure}

The diagram of the effective potential $V_{eff}(r)$
 in Eq. \eqref{71}  for different values of $\gamma$ is depicted in Fig. \eqref{fig:figure 4}.

\subsection{Elapsed time near the singularity}

In this part, we investigate the passage of time near the singularity as it is an important quantity related to metric in Eq. \eqref{5555}. To derive the passage of time, we use  the geodesic equation for a particle that falls radially $u^2=u^3=u^4=0$ as follows

\begin{equation}
		\label{72}
	\frac{du^0}{d\lambda}+\Gamma^0_{\alpha\rho}u^\alpha u^\rho=0.
\end{equation}

After some calculation, we get
\begin{equation}
	\label{73}	
	\frac{d}{d\lambda}\big(g_{00}u^0\big)=0.
\end{equation}

As a result, $g_{00}u^0$  is a constant. Using  normalization condition for massive particles $g_{\alpha\rho}u^\alpha u^\rho=-1$, we have
\begin{equation}
	\label{74}
	u^1=-f^{\frac{1-\mu}{2}}k^\frac{-\nu}{2}\Big(\frac{h^2}{f^\gamma}-1\Big)^{\frac{1}{2}},
\end{equation}

\noindent where $h=g_{00}u^0$. By dividing $u^0$ by $u^1$, we get

\begin{equation}
	\label{75}
	\frac{u^0}{u^1}=\frac{dt}{dr}=-hf^{\frac{\mu-\gamma-1}{2}}k^\frac{\nu}{2}\Big(h^2-f^\gamma\Big)^{-\frac{1}{2}}.
\end{equation}

Near the horizon we assume that $r=\sqrt{2m}+\xi$. By putting this value for $r$ in  \eqref{75} and ignoring the higher order terms, we have ($\theta\neq\pi/2$)

\begin{equation}
	\label{76}
	\frac{dt}{dr}=-\Big(\frac{\sqrt{2m}}{2(r-\sqrt{2m})}\Big)^\frac{\gamma-\mu+1}{2}\Big(\frac{\sin2\theta}{2}\Big)^\nu.
\end{equation}

Therefore the integral of Eq. \eqref{76} yields

\begin{equation}
	\label{77}
	t=\frac{2^{\frac{\mu-\gamma+3}{4}}\big(\sqrt{m}\big)^\frac{\gamma-\mu+1}{2}}{(\gamma-\mu-1)\big(r-\sqrt{2m}\big)^\frac{\gamma-\mu-1}{2}}\Big(\frac{\sin2\theta}{2}\Big)^\nu+Const.
\end{equation}

Also for $\theta=\pi/2$ in Eq. \eqref{75}, we get

\begin{equation}
	\label{78}
	\frac{dt}{dr}=-\Big(\frac{\sqrt{2m}}{2(r-\sqrt{2m})}\Big)^\frac{3\gamma+1}{4}.
\end{equation}

By integrating Eq. \eqref{78}, we obtain the following relations

\begin{eqnarray}
	&&t=\frac{2^{\frac{3}{8}(5-\gamma)}(\sqrt{m})^{\frac{3\gamma+1}{4}}}{3(\gamma-1)\Big(r-\sqrt{2m}\Big)^{\frac{3}{4}(\gamma-1)}}+Const,~~~\gamma\neq 1,\nonumber\\ \\
	&&t=-\sqrt{\frac{m}{2}}\big[\ln(r-\sqrt{2m})\big]+Const,~~~~~~~~~\gamma=1.\nonumber\\
\end{eqnarray}

This result that for case $\gamma=1$ the result reduces to the Schwarzschild case. For an observer at infinite distance from the singularity and for both cases $\gamma=1$ and $\gamma>1$, it takes infinite time for a freely falling particle  to hit the singularity. However, for $\gamma<1$, the observer that is far from the singularity measures a finite time for a freely falling particle to hit the singularity.

\section{Higher DIMENSIONS}

In this section, we introduce two types of three parameter metrics in the presence  
 of a scalar field in flat and toroidal coordinates.

\subsection{Three parameter metrics in toroidal coordinates}

We introduce a new class of three parameter static metrics
 in the presence of a scalar field as follows

\begin{equation}
	\label{90}
	ds^2=-f^\gamma dt^2+\frac{f^\mu}{f}dr^2+r^2f^\mu d\theta^2+r^2f^\frac{1-\gamma}{d-3} \sum_{i=1}^{d-3} d\phi_i^2,
\end{equation}

\noindent where

\begin{equation}
	f(r)=\frac{2m}{r^{d-3}}.
\end{equation}

The scalar field related to metric in Eq. \eqref{90} is 

\begin{equation}
	\Psi(r)=\sqrt{\frac{d-2\gamma-(d-2)\gamma^2}{4(d-3)}-\frac{\mu}{2}}~\ln\Big(\frac{2m}{r^{d-3}}\Big).
\end{equation}

The condition that the scalar field to have real values is as follows

\begin{equation}
	\mu\leq\frac{d-2\gamma-(d-2)\gamma^2}{2(d-3)}.
\end{equation}

The Ricci  scalar for this metric in d-dimensions is as follows

\begin{equation}
	R=\frac{(3-d) m^{1-\mu} \left(\gamma +\frac{1}{2} \gamma ^2 (d-2)+(d-3) \mu -\frac{d}{2}\right) }{2^\mu r^{(d-1)-\mu(d-3)}},
\end{equation}

\noindent and the Kretschmann scalar can be written as below

\begin{eqnarray}
         K&=&\frac{1}{r^{6\mu-2-2d(\mu-1)}2^{2\mu+1}}(d-3)m^{2-2\mu}\nonumber\\
         &\times&\big[2d^2+d-12+4\left(d(d-4)+6\right)\gamma\nonumber\\
         &+&2(d^2(d-3)-1)\gamma^2+4(d(d-3)^2-1)\gamma^3\nonumber\\
         &+&(d-2)(19+2(d-6)d)\gamma^4+4(d-3)(6-3d\nonumber\\
         &+&\gamma-2d\gamma+(5+d-d^2)\gamma^2-(d-4)(d-2)\gamma^3)\mu\nonumber\\
         &+&2(d-3)^2(d-1+4\gamma+2(d-2)\gamma^2)\mu^2\big].
\end{eqnarray}

In this type of coordinates because of the form of the lapse function and the angular coordinates we have only one singularity at $r=0$.

\subsection{Three parameter metrics in flat coordinates}

Another class of three parameter metrics in the presence of a scalar field and in flat coordinates may be written as
\begin{eqnarray}
	ds^2&=&-f^\gamma dt^2+\frac{f^\mu k^\nu}{f}dr^2+r^2f^\mu k^\nu d\theta^2\nonumber\\&+&\theta^2r^2f^\frac{1-\gamma}{d-3} d\phi^2+r^2f^\frac{1-\gamma}{d-3}\sum_{i=1}^{d-4}d\psi_i^2,
\end{eqnarray}

\noindent where

\begin{eqnarray}
	&&\mu+\nu=\frac{1-\gamma}{d-3},\\
	&&f(r)=\frac{2m}{r^{d-3}},\\
	&&k(r,\theta)=\frac{2m}{r^{d-3}}+\frac{(d-3)^2m^2\theta^2}{r^{2(d-3)}},
\end{eqnarray}

\noindent with the following scalar field that is a solution of equation of motion
$(\nabla_\alpha \nabla^\alpha \Theta=0)$

\begin{eqnarray}
	\Theta(r)&=&\sqrt{\frac{(d-2)(1-\gamma^2)}{4(d-3)}-\frac{\nu}{2}}~\ln\Big(\frac{2m}{r^{d-3}}\Big)\\&=&\sqrt{\frac{1}{2}\Big(\mu-\big(\frac{\gamma-1}{2(d-3)}((d-2)\gamma+d-4\big)\Big)}~\ln\Big(\frac{2m}{r^{d-3}}\Big).\nonumber
\end{eqnarray}

The conditions that must be imposed on $\mu$ and $\nu$ so that the scalar field to be real are 

\begin{equation}
	\nu\leq\frac{d-2}{2(d-3)}(1-\gamma^2),
\end{equation}

\noindent and

\begin{equation}
	\mu\geq\Big(\frac{\gamma-1}{2(d-3)}\big((d-2)\gamma+d-4\big)\Big).
\end{equation}

The Ricci and Kretschmann scalars for this metric in d-dimensions are derived as follows

\begin{eqnarray}
	&&R=\frac{2^{\frac{\gamma -1}{d-3}}}{r^{\gamma -(d-3) \nu +(d-2)}} \big[\frac{1}{2} \left(1-\gamma ^2\right) (d-3) (d-2)\nonumber\\&&-(d-3)^2 \nu \big]
	   \left(\frac{1}{2} (d-3)^2 \theta ^2 m+r^{d-3}\right)^{-\nu}m^{\frac{\gamma +d-4}{d-3}},~~~~
\end{eqnarray}
\begin{eqnarray}
	K&=&\Big(2^{\frac{2(\gamma -1)}{d-3}}\frac{(d-3)}{2}\frac{m^\frac{2(d-4)+2\gamma}{d-3}}{r^{(2d-4)+2\gamma-(2d-6)\nu}}\nonumber\\
	&\times&\big(r^{d-3}+\frac{1}{2}(d-3)^2m\theta^2\big)^{-1-2\nu}\Big)\nonumber\\
	&\times&\Big[(d-2)(\gamma+1)^2\big(2d-5+\gamma(2+(2d^2-8d+7)\gamma)\big)\nonumber\\
	&\times&\big(r^{d-3}+\frac{1}{2}(d-3)^2m\theta^2\big)-(4d-12)(\gamma+1)\nonumber
\end{eqnarray}
\begin{eqnarray}
	&\times&\Big(r^{d-3}(2d+\gamma-5)+\frac{m(d-3)^2}{2}\big((2d-5)\nonumber\\
	&+&\gamma(d-1+(d-2)^2\gamma)\big)\theta^2\Big)\nu+(d-3)^3\Big(6r^{d-3}\nonumber\\
	&+&m(d-3)\big((d-1)+4\gamma+2(d-2)\gamma^2)\theta^2\big)\nu^2\Big].~~~~
\end{eqnarray}

It should be noted that due to the form of the lapse function and the angular coordinates we have singularity only at $r=0$.

\section*{conclusion}

Here, we have introduced  a class of three parameter metrics in the presence of a scalar field. We then obtained the effective potential and epicyclic frequencies for stable circular orbit were calculated and their typical behavior were investigated for different values of the parameters that represent the presence of the scalar field. Thereafter, a five dimensional extension of the class of three parameter metrics was introduced and the effective potential and elapsed time near the singularity were also calculated. A lot of properties of the class of four and higher dimensional metrics  can be explored  in the future. The rotating form of this class of metrics is introduced and studied in \cite{Mirza}. Many new aspects of the static and rotating forms of this class of metrics could also be studied in the near future.


\end{document}